# SOLPS-ITER simulation of an X-point radiator in TCV

G. Sun[1], O. Pan[2], M. Bernert[2], M. Carpita[1], C. Colandrea[1], B. P. Duval[1], O. Février[1], S. Gorno[1], E. Huett[1], J.T.W. Koenders[3,4], H. Reimerdes[1], C. Theiler[1], S. Wiesen[5], the TCV team[1]

[1] Ecole Polytechnique Fédérale de Lausanne (EPFL), Swiss Plasma Center (SPC), CH-1015, Lausanne, Switzerland

[2] Max-Planck-Institut für Plasmaphysik, 85748 Garching, Germany

[3] DIFFER-Dutch Institute for Fundamental Energy Research, Eindhoven, Netherlands

[4] Department of Mechanical Engineering, Control Systems Technology Group, Eindhoven University of Technology, Eindhoven, Netherlands

[5] Forschungszentrum Jülich GmbH, Institut für Energie- und Klimaforschung - Plasmaphysik, 52425 Jülich, Germany

SOLPS-ITER simulation is performed to reproduce the X-point radiator recently observed in nitrogen-seeded TCV experiments, which is a scenario that may be favorable to solve the power exhaust problems in future fusion devices. The simulations reveal the transition from the detached regime without XPR to the XPR regime, when increasing the nitrogen seeding rate. A cold X-point core surrounded by ionizing and radiative mentals is formed inside the separatrix and slightly above the X-point, where more than 90% of the total input power is dissipated. The cold X-point core exhibits a temperature of approximately 1eV and features high recombination rate to host the convective fluxes from the ionizing mental. Increasing nitrogen seeding rate also moves the nitrogen ionization front away from the target faster than the nitrogen stagnation point, which enhances the divertor nitrogen leakage to the main chamber and benefits the XPR region cooling. Carbon radiation decreases as the nitrogen seeding increases, and carbon radiation contributes to above 5% of the core impurity radiation before entering the XPR, which decreases to 2.8% when reaching the XPR. Both baffled and unbaffled divertor geometries are simulated and compared, showing that baffles facilitate the access to XPR by increasing the X-point neutral density, but requires higher seeding rate to enter the XPR regime.

## 1. Introduction

Power exhaust remains one of the most critical challenges for future fusion reactors such as DEMO [1, 2]. A stationary and safe divertor operation requires the detached divertor condition which features reduced target particle and power exhaust, and less transient heat pulses from the edge localized modes (ELMs). Recent experiments and simulations in ASDEX Upgrade (AUG) confirmed the existence of a

---

stable X-point radiator (XPR) regime, achieved with strong impurity seeding. A localized, highly radiative region is formed within the closed magnetic flux surfaces slightly above the X-point [3], combined with a detached divertor. ELM suppression was observed in AUG experiments when the XPR moves to a certain height above the X-point [3, 4]. A recent compact radiative divertor based on the XPR concept has been tested in AUG and the divertor remained detached with extremely high input power [5]. The XPR was observed in JET which also has metallic wall as in AUG [6], and in TCV which is equipped with carbon wall [4, 7, 8]. The XPR in TCV was however less stable than in AUG [4], and the underlying mechanisms remain elusive.

The XPR regime has been validated by the numerical simulation code SOLPS 5.0 and SOLPS-ITER for AUG [9-11]. The simulations reproduced a cold X-point core of temperature below 5eV inside the confined region, surrounded by the hotter ionization and radiation mentals, which dissipate the power conducted from upstream and create a high poloidal temperature gradient near the XPR region. A high-recombination region in the cold X-point core was observed, which receives the convective fluxes from the ionization mental. More than 90% of the input power was shown to be dissipated through radiation in the highly radiative XPR region, and both divertors were shown to be detached, consistent with the AUG experiments. These simulations greatly improve the understanding of the XPR regime.

The experimental and numerical XPR results are supported by a recently proposed reduced model for the access to the stable XPR regime and unstable MARFE, based on the particle and power balances [12]. The route to XPR is possible only when the so called stable high-temperature XPR solution is avoided, such that a stable low-temperature XPR solution is achievable. Stable XPR, instead of MARFE, is formed only if the low-temperature XPR solution corresponds to limited recombination rate as it increases exponentially at low temperature levels. The reduced model provides scaling laws which highlights the importance of a range of factors such as divertor neutral density, upstream density, safety factor etc., to access the XPR and prevent the transition to the unstable MARFE. The theory also predicted that impurity-seeded carbon devices are more likely to develop MARFE than in tungsten devices.

To better understand the XPR regime in carbon fusion devices, the SOLPS-ITER code package is employed to reproduce the observed XPR in TCV experiments. The transition from a nitrogen-seeded, detached divertor scenario without XPR, to the nitrogen-seeded XPR regime in single-null (SN) TCV configuration is described, to illustrate the sharp change of particle and power balances when a cold X-point is formed. The neutral and impurity transport in XPR regime is analyzed to help explain the occurrence of a XPR, and the roles of carbon impurity and TCV baffles in XPR formation are discussed. The study aims for further development of the XPR model and support future TCV experiments of XPR, to allow for cross-machine validation of the XPR regime before its implementation in reactor-relevant scenarios.

## 2. TCV experiment and simulation setup

The XPR regime has been achieved in L-mode [8, 13], and H-mode[4] TCV discharges with single-null or snowflake configuration by injecting nitrogen as the extrinsic impurity. The present simulations are based on a typical H-mode XPR discharge with two seeding pulses, Figure 1(a). The discharge features $-210$ kA plasma current, $-1.4$T toroidal field and 1.3MW heating power (NBH). The N II emission peak location was tracked by the multi-spectral imaging diagnostic MANTIS [14], which was used by a real-time controller to adjust the nitrogen seeding rate, and achieve the desired emission peak location during the discharge, Figure 1(e). The XPR windows are defined as the period when the emission peak location is inside the confined region.

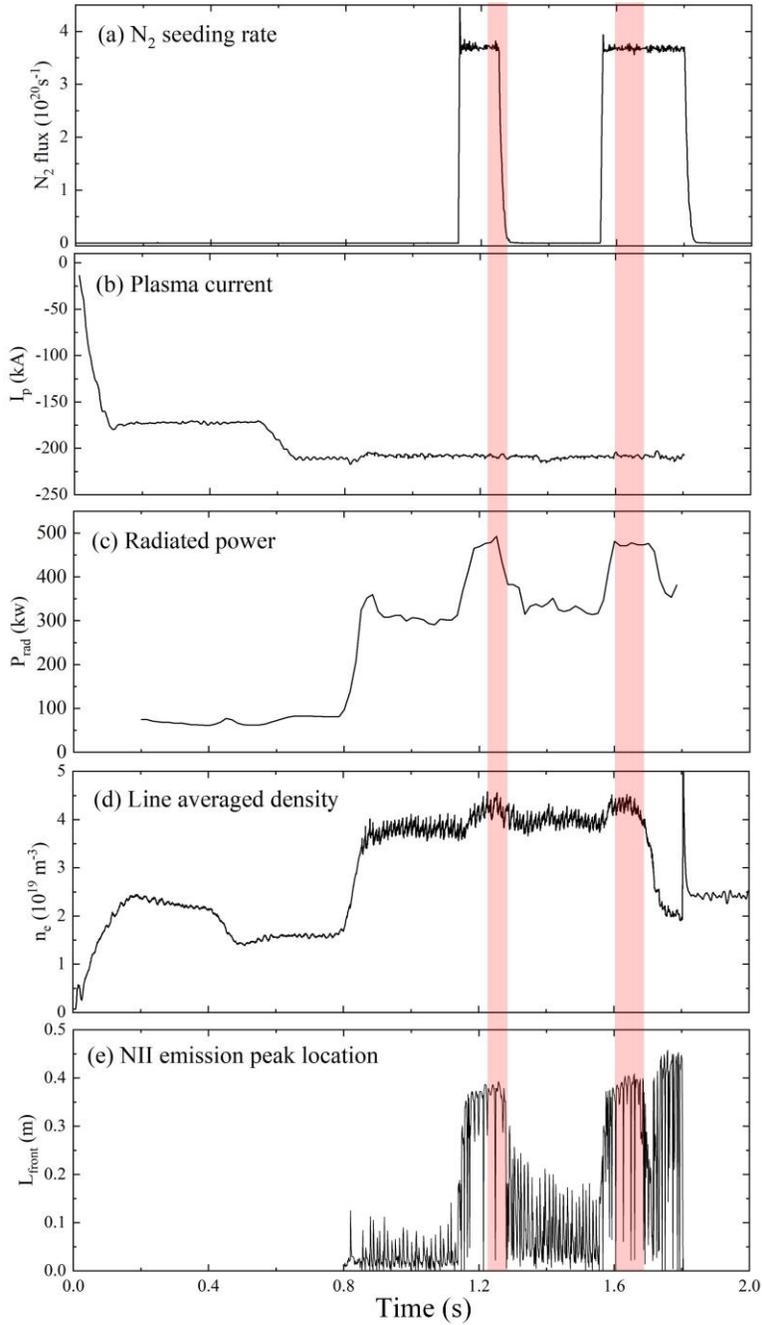

Figure 1. Time traces of XPR discharge 70615 in TCV. (a) Nitrogen seeding rate. (b) Plasma current. (c) Radiated power measured by bolometer. (d) Line averaged density measured by the interferometer along a vertical chord going through the plasma core. (e) Vertical position of the N II emission peak, measured by the MANTIS system, along the outer divertor leg. The target is at 0m and the X-point is at 0.4m. The XPR windows are marked by the shaded areas.

The present simulations are performed using the SOLPS-ITER code package [15]. It consists of the multi-fluid plasma transport code B2.5 and the Monte-Carlo kinetic neutral code EIRENE, and is widely adopted for the scrape-off layer (SOL) plasma studies. The radial limit of the B2.5 grid is mainly restricted by the inner baffle tip. The required magnetic equilibrium for the simulation is chosen at the second XPR window in Figure 1, starting from approximately 1.6s, which is used to generate the simulation grids, Figure 2. Note that both baffled and unbaffled divertor geometries are simulated, which have the same B2 grid but not the same EIRENE grid. The adopted 96 × 36 grid resolution is consistent with previous SOLPS simulations in TCV [16-18], except that the grid resolution near the X-point is increased for the XPR simulation.

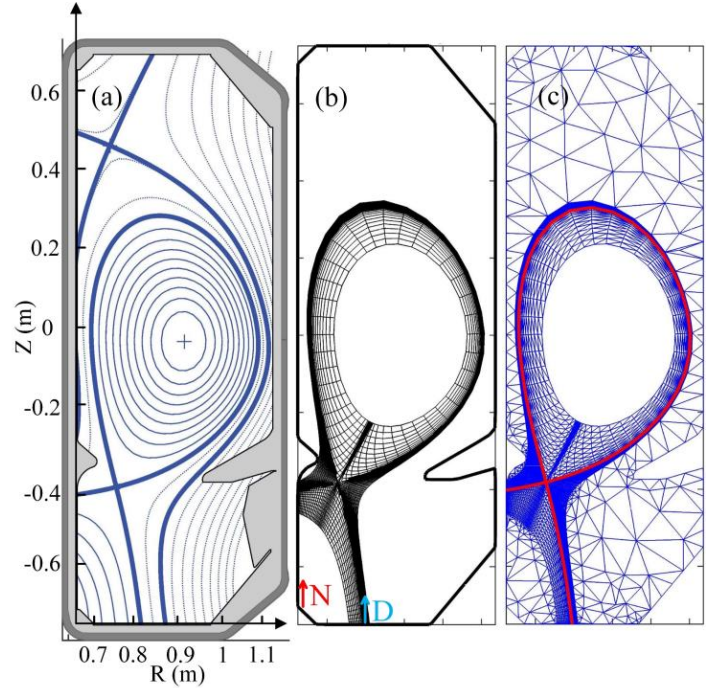

Figure 2. Adopted magnetic equilibrium and the simulation grids. (a) The magnetic equilibrium. (b) B2 grid for plasmas. (c) EIRENE grid for neutrals.

The considered plasma species in the simulations are deuterium, carbon impurities sputtered from the targets and seeded nitrogen impurities. The deuterium and nitrogen are puffed in the common flux region (CFR) and private flux region (PFR), marked in Figure 2(b). Only the atomic nitrogen is seeded in the simulation, assuming short dissociation mean free path of molecular nitrogen. The ammonia formation is neglected. Considered reactions include ionization, charge exchange, dissociation, recombination, elastic collisions and excitation, Table 1. Note that the deuterium-nitrogen charge exchange collision is not yet included in the current version of SOLPS code.

Table 1. Considered reactions

| Reactions |
| --- |
| $D + e \rightarrow D^+ + 2e$ |

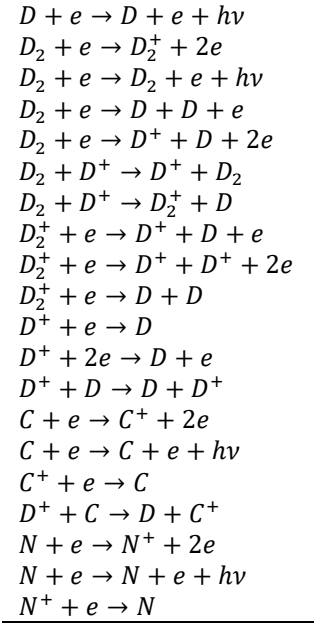

$$D + e \rightarrow D + e + h\nu$$
$$D_2 + e \rightarrow D_2^+ + 2e$$
$$D_2 + e \rightarrow D_2 + e + h\nu$$
$$D_2 + e \rightarrow D + D + e$$
$$D_2 + e \rightarrow D^+ + D + 2e$$
$$D_2 + D^+ \rightarrow D^+ + D_2$$
$$D_2 + D^+ \rightarrow D_2^+ + D$$
$$D_2^+ + e \rightarrow D^+ + D + e$$
$$D_2^+ + e \rightarrow D^+ + D^+ + 2e$$
$$D_2^+ + e \rightarrow D + D$$
$$D^+ + e \rightarrow D$$
$$D^+ + 2e \rightarrow D + e$$
$$D^+ + D \rightarrow D + D^+$$
$$C + e \rightarrow C^+ + 2e$$
$$C + e \rightarrow C + e + h\nu$$
$$C^+ + e \rightarrow C$$
$$D^+ + C \rightarrow D + C^+$$
$$N + e \rightarrow N^+ + 2e$$
$$N + e \rightarrow N + e + h\nu$$
$$N^+ + e \rightarrow N$$

The SOLPS-ITER code does not simulate edge turbulence, and simplifies the anomalous cross-field transport as a radial diffusion dictated by the transport coefficients. The transport coefficients $D_e$, $\chi_e$ and $\chi_i$ are no longer radially constant as in previous L-mode SOLPS simulations in TCV [16, 17, 19, 20]. The transport barrier in H-mode discharges is implemented via certain radial distributions of the transport coefficients. The radial distributions of $D_e$ and $\chi_e$ are determined by fitting with the Thomson scattering (TS) measurements of electron density and temperature at the outer mid-plane (OMP) in upstream, Figure 3. The charge exchange recombination spectroscopy (CXRS) was not activated for the chosen TCV discharge, so the $\chi_i$ fitting against the ion temperature measurement is not performed, and it is assumed that $\chi_e = \chi_i$. After the upstream profile fitting is complete, these transport coefficients are kept constant for all the following simulations.

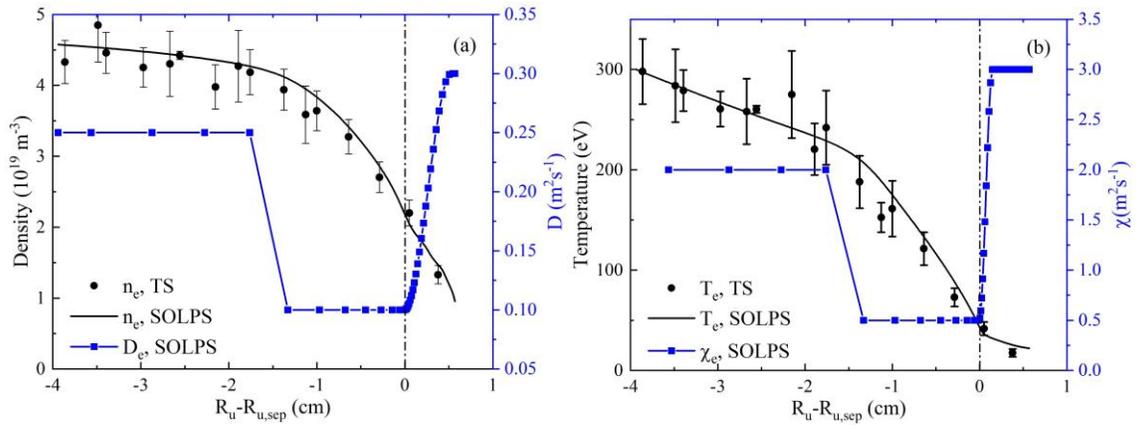

Figure 3. Upstream profiles measured by Thomson scattering and predicted by the SOLPS-ITER simulation, plotted with the transport coefficients $D_e$ and $\chi_e$. (a) Upstream electron density. (b) Upstream electron temperature.

The OMP density is fixed as in Figure 3(a) for different nitrogen seeding rates and divertor geometries throughout the work, by adjusting the deuterium fueling rate accordingly. Fueling rate generally decreases with higher seeding rate and is higher with baffles, for the same upstream density. Recycling coefficients are set to 0.99 for deuterium, 0 for carbon, 1 for neutral and 0.3 for nitrogen ions with a carbon chemical sputtering yield of 3.5%. These values are based on previous simulations of Ohmically-heated TCV plasma discharges that were matched to TCV experimental measurements such as Thomson scattering and divertor spectroscopy [16, 17, 21]. Both the deuterium fueling rate and the nitrogen seeding rate in the simulation are not necessarily consistent with the experimental values, as their exact values depend on the wall recycling rate. Additionally, the seeding rate in TCV experiment is strongly affected by the wall nitrogen storage [4], and the same seeding rate may lead to either over-seeding or is insufficient to enter the XPR due to different discharge history.

The effects of drifts and currents are not considered in the present work. Recent simulation of XPR in AUG with drifts and currents activated suggested improved simulation agreement with the measurement mainly on the high-field side, likely contributed by the radial and poloidal E×B drift [11]. These effects are expected to be explored in our future studies.

## 3. Formation of the XPR and the influence of nitrogen transport

In this section, the transition to the XPR regime is studied by SOLPS-ITER simulation by performing a nitrogen seeding scan. In TCV experiments, the transition to the XPR regime was achieved by increasing the nitrogen seeding rate, until the N II radiation peak moves into the confined region and slightly above the X-point. The nitrogen seeding rate is then actively controlled during the discharge to maintain the radiation peak location. The SOLPS-ITER code solves for the steady-state solution and does not simulate the real-time plasma state evolution, so a constant, feedforward seeding rate is used for each simulation run, with a scan of such seeding rate.

In order to reproduce the XPR initiation, a converged simulation run with fixed core boundary power is first achieved, with the power level consistent with the heating power in TCV experiment. Subsequently, the core boundary condition is changed to fix the electron and ion temperature whose values are the same as the first converged case. The nitrogen seeding rate is then increased until the XPR is observed. The procedure adopted here is to avoid the cold core solution at high nitrogen seeding levels, and is inspired by the previous XPR simulation in AUG [11].

A converged base run with a N seeding rate of $5\times10^{20}$/s is used as a reference case to compare with the XPR run. It has a detached outer divertor but without the XPR. The N seeding rate is then increased until reaching $8.5\times10^{20}$/s, and the XPR is formed. The two representative cases, indexed as the base run and the XPR run (Table 2), are compared and analyzed below to study the XPR characteristics in TCV.

Table 2. Comparison of the base run and XPR run[1].

|  | Base run | XPR run |
|---|---|---|
| Rate ($10^{20}$/s) | 5 | 8.5 |
| $T_{e,XPT}$ (eV) | 28 | 0.9 |
| $T_{e,ot}^{max}$ (eV) | 1.1 | 0.5 |
| $n_{e,ot}^{max}$ ($10^{19}$/m$^3$) | 5.9 | 2.4 |
| $\Gamma_{D+,ot}$ ($10^{21}$/s) | 2.74 | 0.37 |
| $Q_{ot}$ (kW) | 15.8 | 4.3 |

1. Physical quantities from top column to bottom: nitrogen seeding rate, X-point electron temperature, peak outer target electron temperature, peak outer target electron density, outer target D$^+$ flux, outer target heat flux.

The base run and XPR run both have detached outer divertor, and the peak outer target temperature decreases from 1.1eV in the base run to the 0.5eV in the XPR run. The target heat exhaust decreases by 73% from the base run to the XPR run, mainly due to the significant target particle flux reduction, which also leads to a lower carbon sputtering source and will be discussed later. The changes of target profiles are less remarkable than near the X-point, where the X-point temperature decreases from 28eV to 0.9eV. The XPR run also has much colder SOL, and the high-temperature core region is shifted inward, Figure 4. Temperature in OMP separatrix is not affected (not shown).

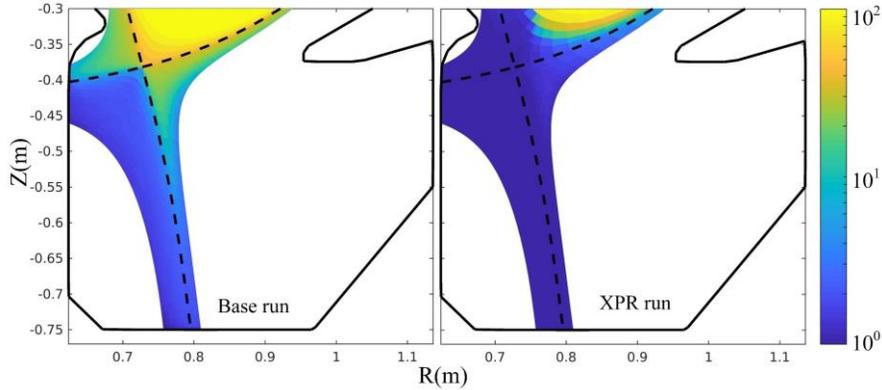

Figure 4. Electron temperature in the X-point and divertor region of the two representative cases.

The cold X-point in XPR run is caused by the strong XPR region power exhaust, contributed by the impurity radiation, charge exchange collision and ionization. In the base run, radiation and ionization are mainly located in the SOL region along the outer divertor leg, and the radiation and ionization fronts are located 0.22m and 0.20m away from the outer target, marked in Figure 5. The front is defined as the poloidal position, between which and the X-point more than 90% of the outer divertor region radiated power/ionization takes place. When the peak radiated power or ionization is inside the confined region as in the XPR run, the front location is set as the X-point. Seeding rate scans of the front locations will be shown later. In the XPR run, both the radiation and ionization peaks inside the confined region, surrounding the cold X-point. The radiation region is located slightly above the ionization region, due to their different temperature dependency and a higher temperature inward. The SOLPS-predicted radiation distribution in the XPR run is consistent with the MANTIS measurement in TCV experiment.

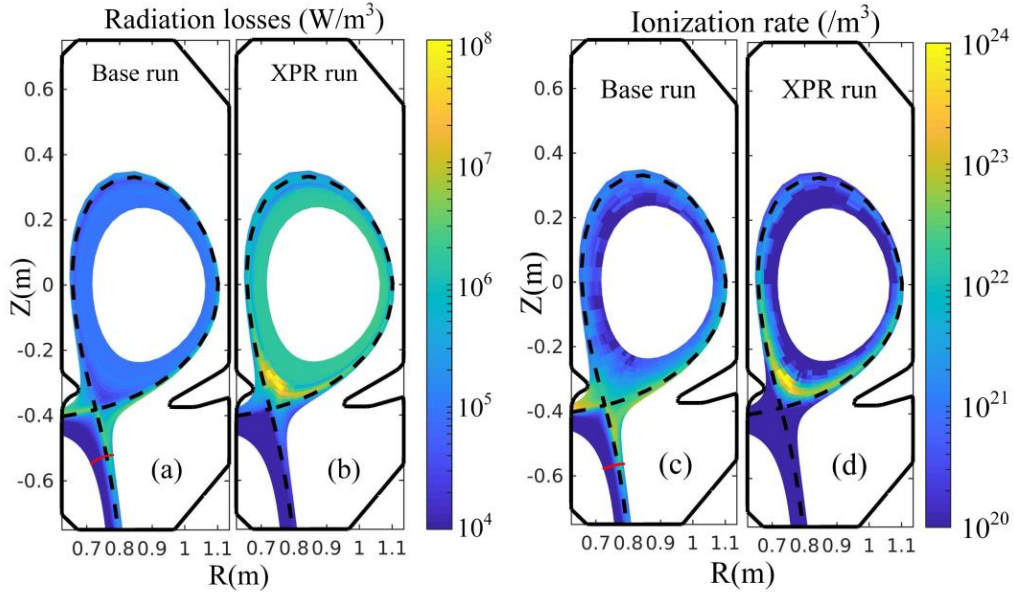

Figure 5. Radiated power density and ionization rate of the two representative cases. Front locations of the base run are marked out.

The recombination region in the base run is nearly invisible compared with the XPR run, and the recombination rate is remarkable only in cells adjacent to the outer target, whose peak temperature is above 1eV. In the XPR run, a strong recombination region is formed near the X-point inside the confined region, as its temperature is below 1eV, Figure 6. The recombination region is located in the cold X-point core, below the strong ionization region and radiation region (simplified as the ionizing and radiative mental below) shown in Figure 5(b)(d). Recombination in the cold X-point core partially supplies the neutrals needed by the ionizing mental above the cold X-point, and hosts part of the convective ion flux, as the recombination rate in the cold X-point core is lower than the ionization rate in the ionizing mental. Most neutrals consumed in the ionizing mental come from the PFR and SOL due to a loss of SOL plasma plugging in the XPR regime.

A region with high neutral density is formed near the inner baffle on the high-field side, compared with the base run, where a stronger neutral leakage to the main chamber region and higher inner mid-plane neutral density are clearly observed, Figure 6(d). The neutral leakage to the main chamber region via the outer baffle and OMP neutral density are less affected in the XPR run compared with the inner baffle, as the cold X-point is located closer to the inner baffle, Figure 4. Whether a longer inner baffle would help to retain the neutral compression in the XPR regime remains to be tested in future works.

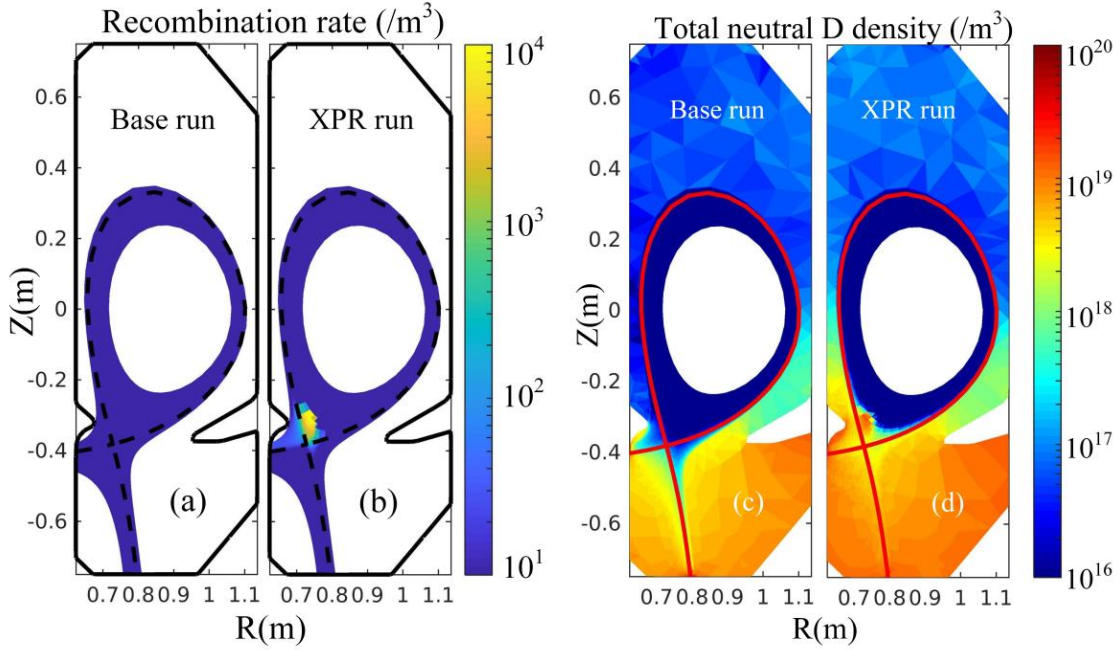

Figure 6. Recombination rate and total neutral deuterium density ($n_{D,tot} = 2n_{D0} + n_D$) of the two representative cases.

Scan of nitrogen seeding rate is performed to better present the quantitative trends of the transition to the XPR regime. The seeding rate ranges between $10^{20}$/s to $9 \times 10^{20}$/s. With higher seeding rate, using the fixed core temperature leads to unphysical solutions and the simulation stops, and using the fixed core boundary power leads to radiative collapse. The XPR window for the nitrogen seeding rate in TCV simulation appears to be narrower than in the AUG simulation [11].

The X-point temperature decreases as the seeding rate increases, until a plateau at approximately 5eV with seeding rate of $7\text{-}8 \times 10^{20}$, Figure 7. In this seeding range, the impurity radiation peak is located at X-point but not yet inside the confined region. A sharp transition to the XPR regime happens only when the seeding rate keeps increasing and the radiation regime moves above the X-point, during which the $T_{e,XPT}$ decreases to approximately 1eV. The transition to the XPR regime is combined with the significant increase of X-point neutral density and reduction of total pressure. The neutral density increases since the plasma plugging near the X-point is weakened, Figure 4. The main sources of the total pressure reduction near the cold X-point core consists of volumetric recombination, which is obvious only when the plasma temperature decreases to sub-eV level, in addition to the charge exchange, cross-field transport, etc.

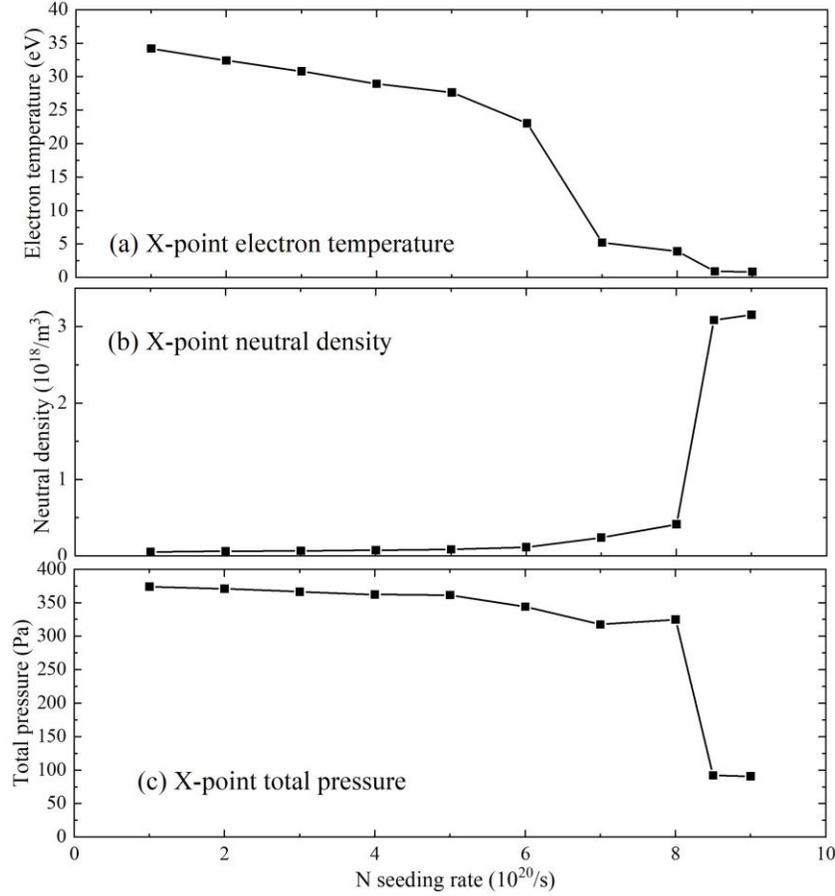

Figure 7. Evolution of the X-point properties with varying nitrogen seeding rate. (a) Electron temperature. (b) Total neutral density. (c) Total pressure.

The transition to the XPR regime is essentially regulated the evolution of nitrogen impurity transport with the increasing seeding rate. When the amount of puffed nitrogen increases from a low level, the radiation in the divertor region increases, which leads to a lower temperature and pushes the deuterium ionization front away from the target. The change of deuterium ionization source modifies the $D^+$ flow pattern, which in turn redistributes the nitrogen impurity transport, via friction force $F_f = m_z(v_{D+} - v_z)/\tau_s$. Here $m_z$ is the mass of impurity ion, $v_{D+}$ and $v_z$ are flow velocities of $D^+$ and impurity ion, and $\tau_s$ is the characteristic stopping time. In a stationary condition as in SOLPS-ITER, the flow rate differences between $D^+$ and impurities are mainly balanced by the thermal force $F_{th} = C_{th}\nabla T$. Here $T$ is the ion temperature and the coefficient $C_{th} \propto Z^2$ with $Z$ the impurity ion charge. Other forces such as pressure gradient force or electrostatic force were shown to be less influential for the impurity transport [22]. The increasing nitrogen seeding rate also alters the poloidal temperature gradient and therefore the thermal force distribution. The impurity stagnation point is defined as the location where $v_z = 0$. Ions above this point are transported upstream and ions below the point are flashed back to the target. At this point the $D^+$ flow velocity satisfies:

$$v_{D+} = -\frac{\tau_s}{m_z} F_{th} \qquad (1)$$

It is however found in the simulation that the stagnation front location only moves slightly upstream from 0.29m to 0.31m, as marked by the blue shaded area in Figure 8(b). This effect is clearly dominated over by the remarkable shift of the neutral nitrogen ionization front upstream shown in Figure 8(b). Here the front is defined as the location between which and the X-point more than 90% of the total nitrogen ionizations in the outer divertor region are included. The definition of the outer divertor region is shown later in Figure 9(c). The ionization front of neutral nitrogen is closer to the outer target with $10^{20}$/s seeding rate, and surpasses the stagnation point with $8 \times 10^{20}$/s seeding rate. With higher seeding rate, the XPR regime is achieved, and the ionization front definition is less meaningful since the majority of the ionization source is located inside the confined region instead of the divertor.

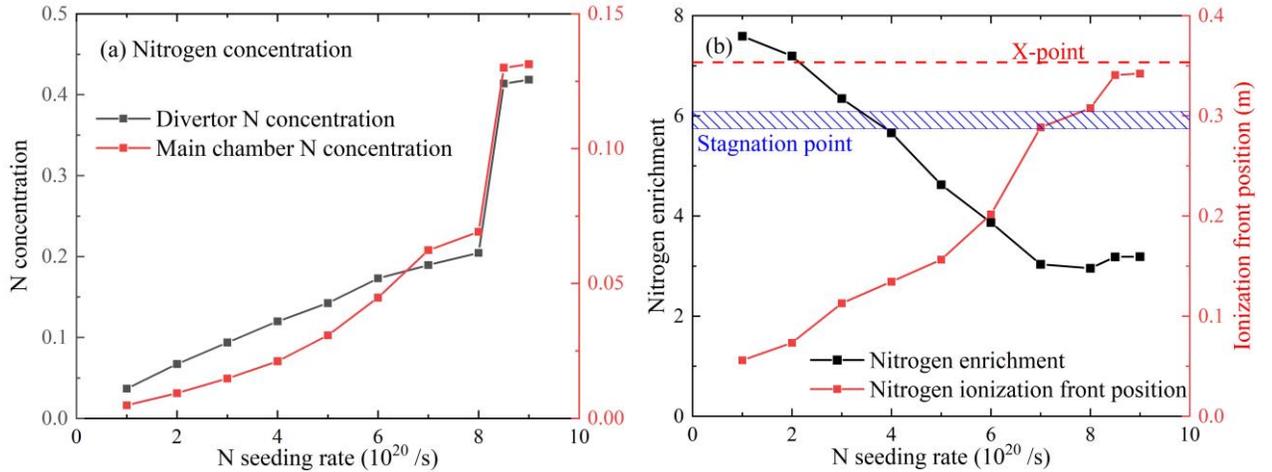

Figure 8. Variation of nitrogen concentration, enrichment, and ionization front poloidal position with the seeding rate. (a) Nitrogen concentration in divertor and main chamber region. (b) Nitrogen enrichment and the neutral nitrogen ionization front poloidal location, counted from the outer target. The X-point location is marked out by a red dashed line and the range of nitrogen stagnation point location is marked by the blue shaded region.

As the neutral nitrogen ionization front moves upstream with increasing seeding rate, higher fractions of the nitrogen neutrals are ionized further upstream than their stagnation point, leading to a higher main chamber region nitrogen concentration, lower divertor region nitrogen concentration, and lower nitrogen enrichment, Figure 8. When sufficient amount of main chamber nitrogen concentration, or impurity radiation in the XPR region, is reached, a cold X-point is formed and the transition to the XPR regime happens, provided that the excessive main chamber region impurity concentration does not lead to the cold core solution (radiative collapse).

The nitrogen concentration is here defined as the sum of all nitrogen ion density normalized by the electron density, averaged in the designated region $c_N = \langle \sum_{i=1}^{7} n_N^{i+} / n_e \rangle$. The main chamber region and divertor region are separated by two connection lines linking the two baffle tips and the X-point. Both diverter ($c_{N,div}$) and main chamber ($c_{N,mc}$) nitrogen concentration increase quasi-linearly with the seeding rate before the XPR threshold seeding rate, and a higher plateau is found in the XPR regime. Note that one concerning issue is that $c_{N,mc}$ in the XPR region appears to be high and can be above 10%, Figure 8(a), which may bring up questions on the core compatibility of the XPR regime in TCV. This concentration is considerably higher than the observed 3% - 4% upstream nitrogen concentration as reported in AUG [11]. The SOLPS-predicted poorer nitrogen retention in TCV might partially explain the experimentally observed narrower operational window for a stable XPR in TCV than in AUG [4]. The nitrogen enrichment is defined as the divertor region concentration divided by the main chamber region concentration $e_N = c_{N,div} / c_{N,mc}$. The enrichment first decreases quasi-linearly with the seeding rate and somehow saturates at approximately 3, for seeding rates above $7 \times 10^{20}$/s.

The nitrogen radiation and density distribution in various regions also vary with the seeding rate, and sharp changes are observed when forming the XPR, Figure 9. Here the nitrogen radiated power and content (ion and neutral number) in a given region divided the total core boundary power and the total nitrogen content are derived, with the region definition shown in Figure 9(c). Nitrogen radiated power fraction in all four regions increases with the seeding rate at relatively low seeding levels until $5\times10^{20}$/s, Figure 9(a), due to higher nitrogen concentration in all regions, Figure 8. The inner and outer divertor radiation then decreases with the seeding rate, when the nitrogen enrichment has almost halved as shown in Figure 8(b). This suggests that the increase of overall nitrogen radiated fraction with the seeding rate is compensated by the enrichment reduction as the ionization front moves upward. Meanwhile, the upper SOL radiation continues to increase with the seeding rate higher than $5\times10^{20}$/s, since the enrichment keeps decreasing. The upper SOL radiation fraction drops sharply when reaching the XPR regime, as the high radiation region moves into the confined region. For similar reasons, the core radiation fraction increases slowly with the seeding rate and rises up sharply when the XPR is formed, reaching 90% of the total core boundary input power. Similar trends are found for the nitrogen content fraction, except that the inner and outer divertor content fraction decreases with the seeding rate at all seeding levels, consistent with the monotonic decrease of the enrichment with increasing seeding rate, Figure 9(b). Above discussed nitrogen retention in the divertor, with respect to the nitrogen leakage to upstream, is crucial for the XPR access. A faster decrease of nitrogen enrichment with increasing seeding rate than in Figure 8(b), may lead to the situation where the core performance has been degraded by the excessive nitrogen impurity before the the peak ionization region moves inside the confined region. In this case, an unstable MARFE instead of stable XPR will happen at high seeding levels. This will be further discussed in section 5 by comparing with the XPR in an unbaffled divertor geometry.

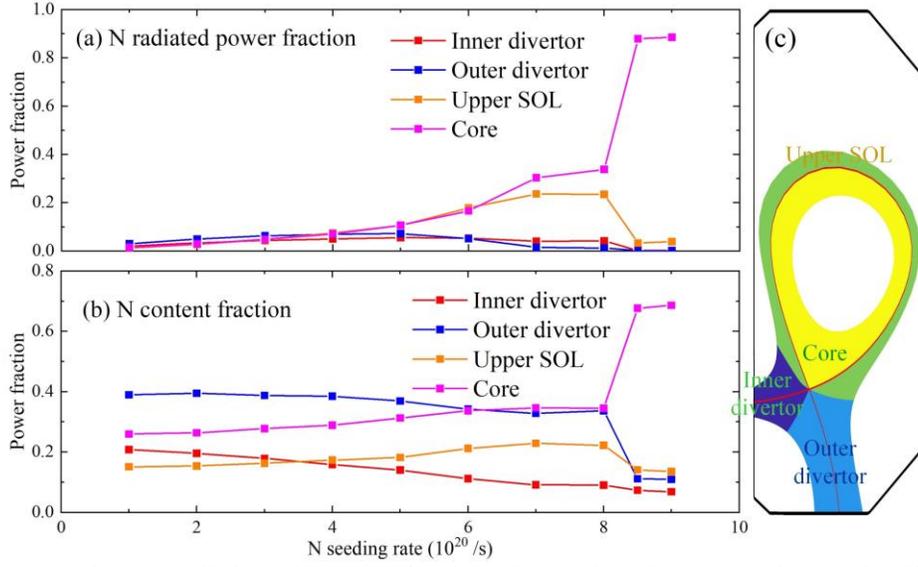

Figure 9. Evolution of the nitrogen radiation and content fraction with varying nitrogen seeding rate in different regions. (a) Regional nitrogen radiated power divided by the total core boundary power. (b) Regional nitrogen ion number divided by the total nitrogen number. (c) Schematic of the region definition.

## 4. Role of carbon impurity

A special condition for the XPR regime in TCV compared with JET and AUG, is that the TCV is equipped with carbon walls, which produce a considerable amount of carbon impurity during the discharge. The existence of carbon impurity, according to the previous XPR theory, is undesirable for the access to XPR [12]. A brief explanation of the theory is given as follows to facilitate the discussion on the role of carbon impurity.

The XPR access theory is based on the particle and power balance of the X-point region. Below it is assumed that the stable high-temperature XPR solution is avoided. Then for the low-temperature XPR solution with a cold X-point core as observed in section 3, the incoming power from the upstream (LHS of the following equation) is fully dissipated by the volumetric power losses (RHS of the following equation) which is mainly the impurity radiation: [12]

$$A_\theta \widehat{k_c} T_u^{7/2} = L_z(T_X) n_X^2 c_{imp,X} V_X \qquad (1)$$

Here $T_X$, $n_X$, $V_X$, $c_{imp,X}$ are mean temperature, density, volume and impurity concentration of the XPR region, $L_z$ is the radiative loss function, $A_\theta$ is the effective XPR poloidal area, $\widehat{k_c} \sim (R_0 q_s)^{-1}$, with $R_0$ the major radius and $q_s$ the safety factor, and $T_u$ is the upstream temperature. The ionization and charge exchange losses are assumed negligible for the low-temperature XPR solution. Solving Equation (1) provides the mean XPR region temperature $T_X$, which is clearly affected the impurity species via $L_z$ and $c_{imp,X}$. The concentration $c_{imp,X}$ was assumed constant in the model. Equation (1) can also be rewritten as follows to facilitate analyses, by assuming pressure balance between the upstream and X-point:

$$\frac{L_z(T_X)}{T_X^2} = \frac{T_u^{3/2}}{n_u^2} \frac{A_\theta \overline{\kappa_c}}{V_X} \frac{4}{c_{imp,X}} \qquad (2)$$

The solved $T_X$ must be judged by another criterion derived from the particle balance to prevent the XPR from evolving into an unstable MARFE, with $\Gamma_{rec} > \Gamma_{in}$ signifying a transition to MARFE, meaning the poloidal particle flux entering the XPR region is less than the volumetric recombination particle loss in the XPR region. This criterion is independent of the impurity species. The calculation results suggest that pure carbon radiation leads to a much narrower upstream density window for XPR without entering MARFE, than nitrogen and argon. Investigating the fraction of carbon radiation in the XPR region is hence critical to understand whether the carbon wall in TCV hinders the XPR access.

SOLPS simulations suggest that the fraction of carbon core radiated power, normalized by the total core radiated power (indexed as $f_C$), is higher than the nitrogen fraction $f_N$ with $10^{20}$/s nitrogen seeding rate. $f_C$ decreases and $f_N$ increases with increasing seeding rate. $f_C$ is below 3% even before entering the XPR regime, and further decreases to 2.8% in the XPR regime. The decreases of $f_C$ is explained by the reduced carbon sputtering source, which is essentially due to less ionization source and hence lower target particle flux, at high seeding levels [17].

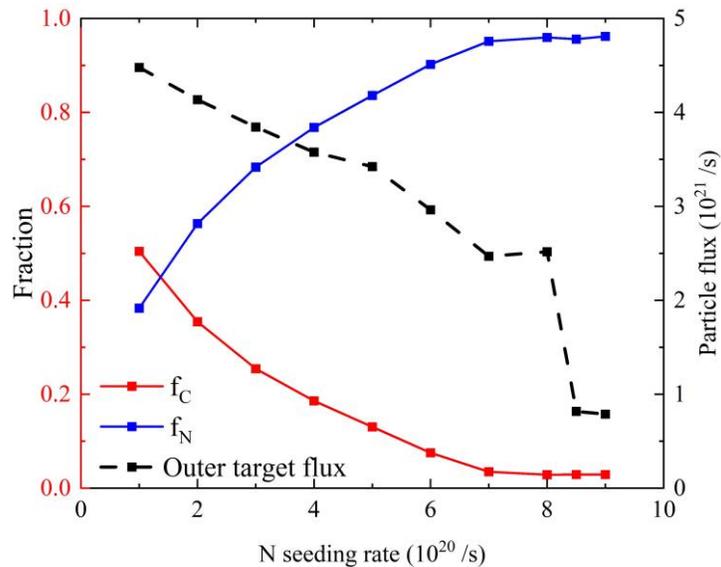

Figure 10. Core radiated power of carbon and nitrogen, normalized by the total core radiated power, and the outer target particle flux, with varying nitrogen seeding levels.

Though the carbon radiation only accounts for a limited fraction of radiated power at high seeding levels, it might strongly affect the XPR to MARFE transition in real TCV discharges. The SOLPS-ITER is a steady-state simulation code and cannot reveal the real-time evolution when more seeded nitrogen brings down the $T_X$ to a low level. During this process, the presence of carbon can destabilize the XPR when $T_X$ approaches approximately 10eV, where the radiation efficiencies of carbon and nitrogen peak, and are orders of magnitude higher than at surrounding temperatures. If at this $T_X$ level, carbon density is still relatively high, the drastically different $L_z$ for carbon and nitrogen would split the radiation peak and shift the low temperature XPR solution into the unstable MARFE.

The present simulations give a considerable carbon radiation fraction $f_C$ of approximately 5% at the threshold temperature 10eV, assuming linear interpolation of physical quantities between the discrete nitrogen seeding level. Such level of carbon radiation should be sufficient to contribute to the experimentally observed less stable XPR in TCV than in metallic-wall machines. More detailed analyses are expected to better explain the role of carbon impurity in XPR access.

## 5. Influence of neutral baffling

It has been proved numerically and experimentally that the TCV baffles can significantly increase the neutral pressure in the divertor region [16, 18, 23, 24]. The XPR access model introduced in section 4 pointed out the important of the neutral pressure in the XPR region, which is just below the TCV baffles and should benefit from the improved divertor neutral pressure. An explicit scaling law of the XPR access condition, derived by avoiding the stable high-temperature XPR solution using the XPR power balance, is as follows:[12]

$$X_A \sim \frac{R_0^2 q_s^2 f_{exp}}{a} \frac{n_{0X} n_u}{T_u^{2.5}}$$ (3)

The XPR access factor $X_A > 1$ is required to avoid the high-temperature solution, which means that the XPR ionization and charge exchange losses exceed the conductive power towards the XPR region. Impurity radiation is assumed to be negligible in the considered temperature range. $a$ is the minor radius, $f_{exp}$ is the upstream-to-XPR flux expansion, $n_{0X}$ is the mean neutral density in the XPR region and $n_u$ is the upstream density. In the present simulations, the upstream quantities are controlled, and the flux expansion is shown to vary little with baffles, hence the access factor $X_A$ mainly depends on $n_{0X}$.

Scan of nitrogen seeding rate is performed for the unbaffled divertor, same as the baffled divertor, for comparison. The seeding scan with unbaffled divertor stops at a lower seeding level than the baffled divertor, at $6 \times 10^{20}$/s, since the low temperature XPR solution in the unbaffled divertor can be achieved with less amount of nitrogen. This is likely due to a higher impurity neutral conductance from the divertor to the main chamber region upstream and lower nitrogen enrichment, Figure 11. It leads to higher nitrogen concentration in the main chamber region, which radiates more power and reduces the amount of conductive power entering in the XPR region, hence making an unbaffled divertor easier to achieve low $T_X$ and the XPR than the baffled divertor, with the same nitrogen seeding rate and upstream conditions.

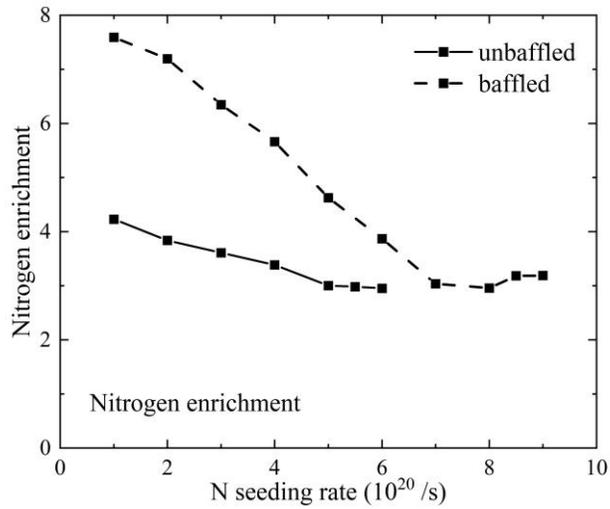

Figure 11. Nitrogen enrichment with unbaffled and baffled divertors, with varying nitrogen seeding levels.

The unbaffled neutral density at the X-point is lower than half of the baffled X-point neutral density, also the neutral compression is much lower without baffles, Figure 12. Here the neutral compression is defined as the mean neutral density in the divertor region and in the main chamber region. The two regions are separated by two segments connecting the baffle tips and the X-point. The baffles increases the compression by decreasing the divertor-to-main chamber neutral conductance, and increasing the fraction of recycled neutrals which are not immediately ionized in front the target .

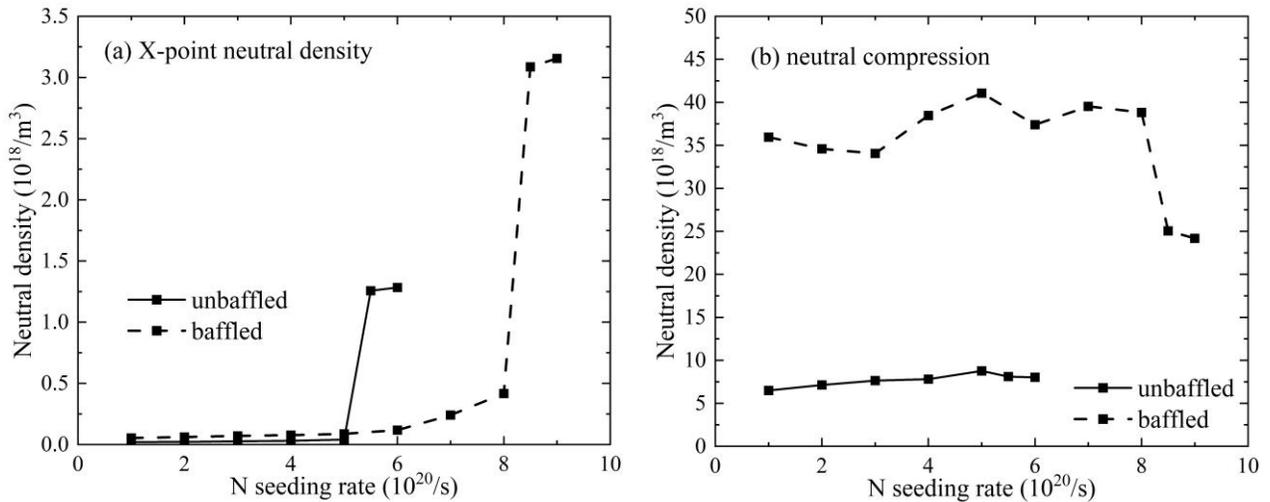

Figure 12. Comparison of (a) X-point neutral density and (b) neutral compression with unbaffled and baffled divertors, with varying nitrogen seeding levels.

In summary, the effects of baffles on XPR are twofold: increasing the XPR region neutral density, and decreasing the main chamber region nitrogen density. According to the XPR theory [12], the former helps to evade the stable high-temperature XPR solution, which enables the route to the XPR solution and broadens the XPR window. The latter requires higher nitrogen seeding rate to achieve the low-temperature XPR solution but has no obvious effect on the transition to unstable MARFE. Simulations of XPR with different baffle shapes will be performed in future works.

## 6. Conclusions

The X-point radiator regime observed in recent TCV experiments is reproduced with the SOLPS-ITER simulation in H-mode, SN configuration with nitrogen seeding and baffled divertor. When increasing the seeding rate to a sufficiently high level, a transition from a regime with detached divertor and hot X-point to the XPR regime with cold X-point is observed. The radiation, ionization sources are mainly located in a region slightly above the X-point, combined with strong volumetric recombination near the cold X-point. Both the radiative and ionizing mentals contribute to a strong poloidal temperature gradient and form a cold X-point core. The nitrogen ionization front is found to move away from the target with increasing seeding rate, and approaches the X-point before entering the XPR regime. The movement of nitrogen ionization front, instead of the movement of nitrogen stagnation front, dominates the nitrogen transport. The carbon concentration is found to decrease with increasing nitrogen seeding rate, due to less sputtering sources with the target current roll-over. The carbon concentration is however nonnegligible before entering the XPR, which may contribute to the less stable XPR in carbon device than in tungsten device observed in experiment. The influence of TCV baffles on XPR access is proved to be twofold. A higher X-point neutral density due to baffling broadens the XPR operation window by enhancing the XPR region ionization losses, but higher nitrogen seeding rate is required to achieve the XPR.

## ACKNOWLEDGMENTS

This work has been carried out within the framework of the EUROfusion Consortium, via the Euratom Research and Training Programme (Grant Agreement No. 101052200— EUROfusion) and funded by the Swiss State Secretariat for Education, Research and Innovation (SERI). Views and opinions expressed are however those of the author(s) only and do not necessarily reflect those of the European Union, the European Commission, or SERI. Neither the European Union nor the European Commission nor SERI can be held responsible for them. This work was supported in part by the Swiss National Science Foundation.